# The Effects of Varying Penetration Rates of L4-L5 Autonomous Vehicles on Fuel Efficiency and Mobility of Traffic Networks

Ozgenur Kavas-Torris, M. Ridvan Cantas, Karina Meneses Cime, Bilin Aksun-Guvenc, Levent Guvenc

Automated Driving Lab, Ohio State University

## Abstract

With the current drive of automotive and technology companies towards producing vehicles with higher levels of autonomy, it is inevitable that there will be an increasing number of SAE level L4-L5 autonomous vehicles (AVs) on roadways in the near future. Microscopic traffic simulators that simulate realistic traffic flow are crucial in studying, understanding and evaluating the fuel usage and mobility effects of having a higher number of autonomous vehicles (AVs) in traffic under realistic mixed traffic conditions including both autonomous and non-autonomous vehicles. In this paper, L4-L5 AVs with varying penetration rates in total traffic flow were simulated using the microscopic traffic simulator Vissim on urban, mixed and freeway roadways. The roadways used in these simulations were replicas of real roadways in and around Columbus, Ohio, including an AV shuttle routes in operation. The road-specific information regarding each roadway, such as the number of traffic lights and positions, number of STOP signs and positions, and speed limits, were gathered using OpenStreetMap with SUMO. In simulating L4-L5 AVs, the All-Knowing CoEXist AV and a vehicle with Wiedemann 74 driver were taken to represent AV and non-AV driving, respectively. Then, the driving behaviors, such as headway time and car following, desired acceleration and deceleration profiles of AVs, and the non-AVs' car following and lane change models were modified. The effect of having varying penetration rates of L4-L5 AVs were then evaluated using criteria such as average fuel consumption, existence of queues and their average/maximum length, total number of vehicles in the simulation, average delay experience by all vehicles, total number of stops experienced by all vehicles, and total emission of CO, NOx and volatile organic compounds (VOC) from the vehicles in the simulation. The results show that while increasing penetration rates of L4-L5 AVs generally improve overall fuel efficiency and mobility of the traffic network, there were also cases when the opposite trend was observed.

## Introduction

Since Autonomous Vehicles (AV) behave differently than human drivers, it is valuable to study the possible effects of SAE automation level L4-L5 AVs on fuel economy, mobility and emissions in a traffic simulation environment. By doing so, the automotive industry and departments of transportation can prepare for the possible changes and challenges of having L4-L5 AVs on roadways in the near future.

There is on-going research about the effects of having AVs in traffic and some of the available results will be discussed next. It was presented in [1] that one AV is able to regulate the flow of up to 20 vehicles around it and can help in the dissipation of stop-and-go waves. In [1], an actual vehicle with AV capabilities was used during the experiments with multiple conventional non-AV vehicles. Since it is very difficult to run such actual vehicle tests on public roads in controlled test cases, the use of traffic simulators becomes crucial in simulating AVs and showing their effects on the non-AVs in traffic.

There is a considerable number of studies that were carried out by focusing on simulating AVs in traffic networks in the literature. For instance, Lackey in [2] simulated L2, L3 and L4 AVs in the SUMO Microscopic Traffic Simulator on three different urban and highway routes to investigate how the AVs affected the overall roadway mobility in a traffic network. He concluded that the frequent lane changing behaviors of AVs improves traffic mobility significantly. Wang [3] tested an external longitudinal driving model they developed in Vissim and concluded that using their longitudinal model resulted in a 1.13% increase in the number of vehicles per hour in the traffic network over conventional human drivers. Shladover et. al [4] showed that vehicles simulated in the AIMSUN microscopic simulator that were equipped with a Cooperative Adaptive Cruise Control (CACC) system could improve the freeway lane effective capacity. Morando et. al. [5] investigated the safety impacts of AVs around a signalized intersection and a roundabout in Vissim and concluded that AVs would improve safety significantly. Stanek et. al. [6] simulated AVs in Vissim and studied the effects of AVs on congested networks. Olia et. al [7] studied the effects of automated vehicles mixed with regular vehicles in highway systems. Talebpour et. al. [8] worked on how the Autonomous and Connected vehicles for different market penetration rates affect the stability of traffic flow and concluded that automation could prevent shockwave formation and propagation in traffic flow. Aria et. al. [9] studied how the automated vehicles affect conventional driver behavior and their traffic performance with microscopic traffic simulations and concluded that AVs helped especially at peak hours. Litman [10] studied the predictions about AV implementation implications for transport planning and concluded that traffic, parking congestion, energy conservation, safety and pollution reductions are some of the main areas that AV are expected to positively affect in the next decade.

This paper focuses on simulating L4-L5 AVs in the Vissim traffic simulator to study the effects of varying degrees of presence of AVs on fuel economy, mobility, queue length, vehicle delay and vehicle emissions. For each route, a short 500 seconds simulation and a long 3,600 seconds simulation were run for different degrees of AV penetration rate, meaning 0% AVs, 20% AVs, 35% AVs, 50% AVs, 65% AVs, 80% AVs and 100% AVs in the traffic network. These simulations were repeated for several different routes and simulation results were used to understand the effect of AV penetration rate on fuel economy and mobility.

The remainder of the paper is organized as follows: Simulation of L4-L5 AVs in a Traffic Network section explains how the AVs were simulated in Vissim for this study. Traffic Network Simulation Results section presents the results of the study for four different routes around Columbus, Ohio. Conclusions and Future Work section summarizes the work done, draws conclusions about the study and elaborates on how to extend this study for future work.

## Simulation of L4-L5 AVs in a Traffic Network

Various ways are possible to design and simulate L4-L5 autonomous vehicles in a traffic environment. The traffic environment chosen for this study was a traffic simulator called Vissim. Using Vissim, it is possible to simulate AVs and see the interaction between conventional vehicles and AVs.

The All-Knowing CoEXist Autonomous Vehicle model in [7] that can be found in the microscopic traffic simulator Vissim is a good starting point in simulating autonomous vehicles. The All-Knowing CoEXist AV had been modeled, as explained in [7], using the field data from a test track in Helmond, Netherlands.

Taking the All-Knowing CoEXist AV as a baseline, the All-Knowing CoEXist AV was duplicated to be modified in the Vissim environment. as far as the definition of the L4-L5 AV used in this study, the AVs studied in this work are L4-L5 level AVs equipped with CAV functionalities, such as V2V and V2I communication.

Additionally, by changing vehicle input configurations, the number of L4-L5 AVs in the full vehicle network was varied to see the effect of L4-L5 AVs in a traffic network. According to predictions about how the AVs will behave in the future, the desired acceleration and deceleration profiles of the All-Knowing CoEXist AVs were also modified.

For the acceleration and deceleration profile, as well as desired speed profile modifications, guidelines given in [7] were used. For instance, automated vehicles are expected to behave more deterministically instead of stochastically. With this assumption, the acceleration and deceleration profiles were changed to bring the max and min curves given in Vissim closer to the median curve. As seen in Figure 1 on the left side, the original conventional vehicle model in Vissim had the given maximum acceleration profile with maximum and minimum curves (shown in green), as well as the median (red dotted line). By getting the maximum and minimum curves as close to the median as possible, the expected deterministic behavior of L4-L5 AV maximum acceleration behavior was achieved (Figure 2 on the right side).

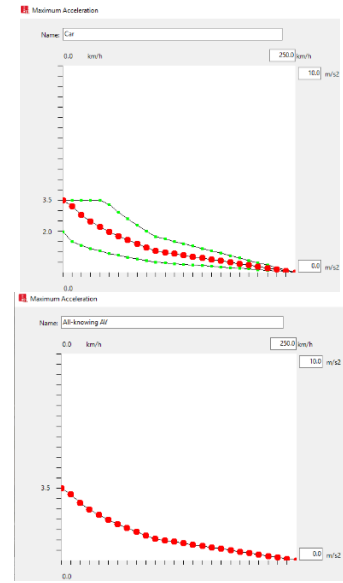

Figure 1: Maximum Acceleration Behavior Modification for L4-L5 AV

The number of vehicles or objects that the AVs interact with were also increased to 10 objects. Another variable that was modified was the car following behavior of the CoEXist vehicle. For the car following, the time-gap for the AV was decreased, since AVs will be equipped to follow vehicles in front of them more closely, leaving a smaller time gap between themselves and their preceding vehicles.

Another aspect of the lane changing and car following behavior came about because of the ability that AVs have to move cooperatively and communicate with each other, as well as to interact with non-AVs and traffic infrastructure around them. Thanks to V2V, V2X and V2I communication abilities of the L4-L5 AVs, they are able to change lanes more frequently and suddenly without having unnecessary braking. This elimination of unnecessary braking is expected to result in an increase in fuel economy benefit with higher penetration rates of AVs.

The modified L4-L5 AVs in Vissim can be seen in Figure 2. In Figure 2, the bright pink geometric stadium represents the L4-L5 AV and the other geometric stadiums represent the non-AVs in the traffic simulation.

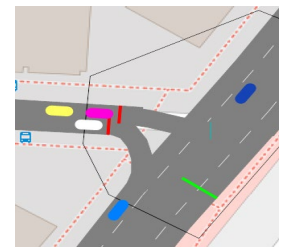

Figure 2: AVs and non-AVs in Traffic

For simulating the non-AV vehicles in the traffic network, conventional vehicles with the Wiedemann 74 [11] driver model were used. Taking that model as the starting point, the car following and lane changing behavior of the Wiedemann 74 driver was modified to better suit the behavior of a conventional driver. The number of interaction vehicles around non-AVs were modified in the Wiedemann 74 driver model since conventional vehicles do not interact with as many number of cars or objects around them as AVs do. The acceleration and deceleration profiles of the Wiedemann 74 driver were also modified.

Some assumptions were made for the modifications on the acceleration and deceleration behaviors of the vehicles in the traffic simulation. The non-AVs are expected to accelerate and decelerate more slowly than L4-L5 AVs. Non-AVs are also expected to go slower than L4-L5 AVs in traffic. Based on these expectations, the necessary modifications were carried out in the acceleration and deceleration profiles of L4-L5 AVs and non-AVs.

Several test routes were designed and tested to observe the effects of L4-L5 AVs in a traffic network. The routes tested in this study can be seen in Table 1. These routes are shown graphically in the corresponding simulations in the following Traffic Network Simulation Results section. Each scenario for each route was tested with 10 runs using the random seed capability of Vissim. Then, averages of each run were recorded to be presented.

Table 1: Routes tested in the study

| Route Name | Traffic Condition |
|---|---|
| Route 19 | Urban + Freeway |
| Route 15 | Urban only |
| Route US 33 | Freeway only |
| Route COSI | Urban only |

Information about Routes 19, 15 and US 33, such as the total length, number of traffic lights and signal phasing and timing (SPaT) for each traffic light, was gathered by importing OpenStreetMap maps into SUMO and running simulations with SUMO while using Traci [12]. The necessary information was then exported from the route and the routes were modeled in Vissim.

For the determination of traffic volume for each route, THE Ohio Department of Transportation Data Management System was used [13]. In their website, the roads that different networks consisted of were found and the data regarding the traffic volume for said roads were implemented in the simulation environment.

## Traffic Network Simulation Results

After the simulation environment for each route was generated in Vissim, 10 simulations were run for each AV penetration rate. The first simulation was run around an intersection that was part of the route being examined and the simulation time was 500 seconds. The simulation run breakdown can be seen in Table 2.

Table 2: Simulation Breakdown

| AV Penetration Rate | Runs per Route | | Total Number of Simulation Runs for 4 Routes |
|---|---|---|---|
| | For 500 seconds | For 3600 seconds | |
| No AV | 10 | 10 | 700 |
| 20% AV | 10 | 10 | |
| 35% AV | 10 | 10 | |
| 50% AV | 10 | 10 | |
| 65% AV | 10 | 10 | |
| 80% AV | 10 | 10 | |
| 100% AV | 10 | 10 | |
| Total number of simulations per Route | 140 | | |

The second simulation was run for the whole traffic network with multiple intersections. For the second set of simulations, a longer simulation duration was chosen and the simulation duration was set as 3,600 seconds (1 hour).

In the following subsections, the fuel economy benefit and mobility effects of the AVs are presented for Route 19, Route 15, Route US 33 and Route COSI, respectively. Then, the last subsection in this Traffic Network Simulation Results section quantifies the effects of having AVs using metrics such as queue length, vehicle emissions and vehicle delay.

### Traffic Network Route 19

The effects of having AVs in traffic was studied around an intersection first (Figure 3). This intersection is part of a whole traffic network that spans urban and highway roads in a route called Route 19.

In order to regulate the traffic around the intersection shown in Figure 3, traffic lights with light durations similar to real life traffic lights were also modeled and can be seen as the red bars in Figure 3.

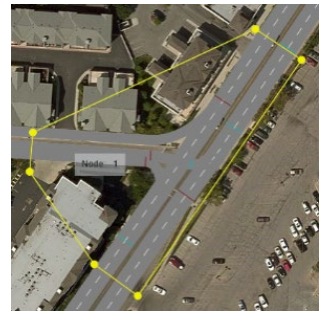

Figure 3: Route 19 Intersection in Vissim

Route 19 can be seen in the map of Figure 4. It is the replica of a real route in Columbus, Ohio. The total length of Route 19 is 7 km and it has 5 traffic lights along the route.

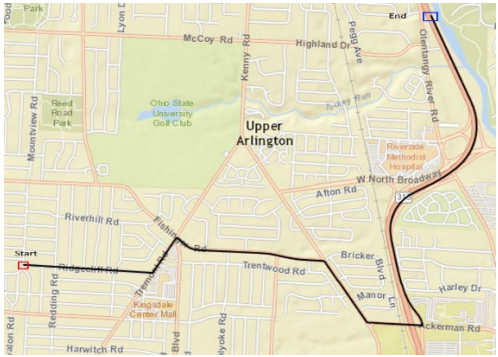

Figure 4: Route 19

Then, the simulation results were summarized in showing fuel efficiencies and total number of vehicles around the intersection (Table 1). Looking at Table 1, it can be seen that for penetration rates of AVs in the traffic network from 0% to 20%, the percentage fuel economy benefit increases with increasing AV penetration. From 20% to 35% penetration of AVs, the fuel economy benefit decreased. From 35% to 100% penetration of AVs, the fuel economy benefit increased and reached a maximum of 14.6% at a penetration rate of 100% L4-L5 AVs in the traffic network. It is also seen that when the L4-L5 AV penetration rate increases in a traffic network, since the vehicles move smarter, they can move at a higher speed, resulting in having a higher number of vehicles in the traffic network and having increased mobility.

Table 3: Route 19, Intersection Results

| Route 19, Intersection Results – Urban + Freeway Traffic | | | | | |
|---|---|---|---|---|---|
| Simulation Duration in seconds | AV Penetration Rate | Number of Vehicles | Total Fuel Consumption in US gallons | Fuel Consumption per Vehicle in US gallons | % Fuel Economy wrt no AV case |
| 500 seconds | No AV | 162 | 1.85383 | 0.01144 | 0.000 |
| | 20% AV | 165 | 1.85720 | 0.01129 | 1.279 |
| | 35% AV | 166 | 1.88086 | 0.01133 | 0.928 |
| | 50% AV | 168 | 1.86087 | 0.01108 | 3.186 |
| | 65% AV | 170 | 1.86702 | 0.01102 | 3.692 |
| | 80% AV | 172 | 1.79269 | 0.01045 | 8.644 |
| | 100% AV | 175 | 1.70830 | 0.00977 | 14.628 |

The summarized results can be seen in Figure 5. As the penetration rate of L4-L5 AVs increased in the traffic network, the number of vehicles going in and out of the intersection increased as well. The number of vehicles around the intersection increased considerably and reached a maximum of 8% compared to the no-AV case at 100% AV penetration rate.

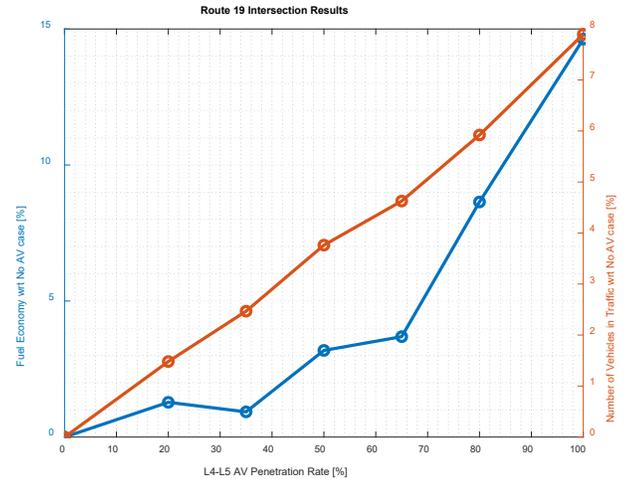

Figure 5: Fuel Economy and Number of Vehicles in Traffic Network vs L4-L5 AV Penetration Rate for an Intersection in Route 19

For the whole Route 19, the results can be seen in Table 4. For the 3,600 seconds simulation, the mobility and fuel economy both showed an increasing trend with respect to increasing L4-L5 AV penetration rates, except at 65% AV penetration rate.

Table 4: Route 19, Full Route Results

| Route 19, Full Route Results – Urban + Freeway Traffic | | | | | |
|---|---|---|---|---|---|
| Simulation Duration in seconds | AV Penetration Rate | Number of Vehicles | Total Fuel Consumption in US gallons | Fuel Consumption per Vehicle in US gallons | % Fuel Economy wrt no AV case |
| 3.600 seconds | No AV | 3996 | 372.55360 | 0.09321 | 0.000 |
| | 20% AV | 4196 | 362.30330 | 0.08631 | 7.408 |
| | 35% AV | 4317 | 360.47950 | 0.08348 | 10.437 |
| | 50% AV | 4466 | 369.35350 | 0.08270 | 11.279 |
| | 65% AV | 4665 | 387.17020 | 0.08300 | 10.957 |
| | 80% AV | 4930 | 381.16070 | 0.07731 | 17.059 |
| | 100% AV | 5027 | 358.17840 | 0.07124 | 23.578 |

The summarized results for the 3,600 seconds simulation on Route 19 can be seen in Figure 6. From the No AV case to penetration rate of 100% L4-L5 AV case, the mobility of the traffic network increased by about 28%, while the fuel economy benefit increased by 23.6%.

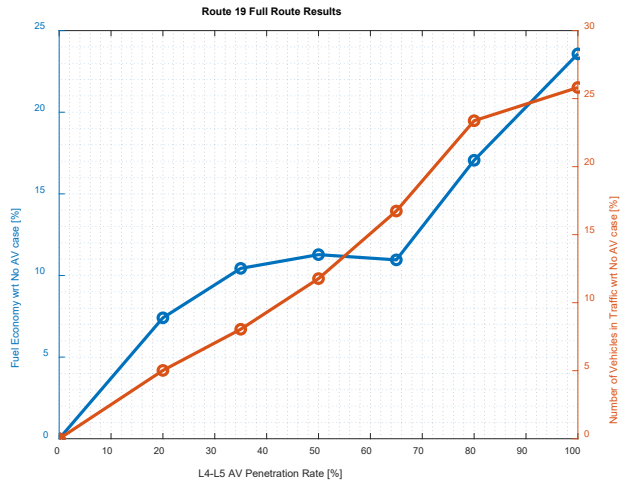

Figure 6: Fuel Economy and Number of Vehicles in Traffic Network vs L4-L5 AV Penetration Rate for Route 19

### Traffic Network Route 15

For Route 15, the intersection studied can be seen in Figure 7. This intersection is part of a whole traffic network that spans an urban only route.

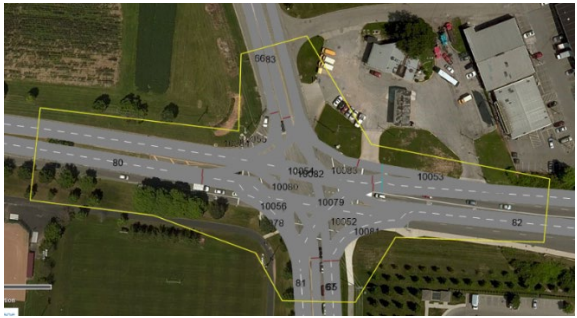

Figure 7: Route 15 Intersection in Vissim

Route 15 can be seen in the map of Figure 8. It is the replica of a real route in Columbus, Ohio. The total length of Route 15 is 7.4 km and it has 14 traffic lights along the route.

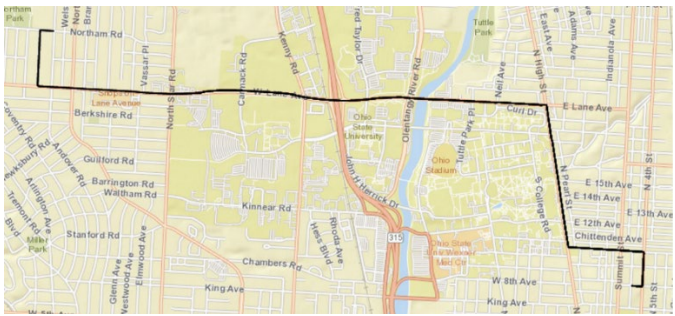

Figure 8: Route 15

For the intersection shown in Figure 7 on Route 15, the simulation results can be seen in Table 5. It is observed that having L4-L5 AVs in the traffic network resulted in increased mobility for each penetration rate.

Table 5: Route 15, Intersection Results

| Route 15, Intersection Results – Urban Only Traffic | | | | | |
|---|---|---|---|---|---|
| Simulation Duration in seconds | AV Penetration Rate | Number of Vehicles | Total Fuel Consumption in US gallons | Fuel Consumption per Vehicle in US gallons | % Fuel Economy wrt no AV case |
| 500 seconds | No AV | 84 | 1.00240 | 0.01190 | 0.000 |
| | 20% AV | 86 | 1.00240 | 0.01166 | 1.962 |
| | 35% AV | 87 | 1.03420 | 0.01183 | 0.527 |
| | 50% AV | 89 | 1.05790 | 0.01182 | 0.642 |
| | 65% AV | 91 | 1.07430 | 0.01182 | 0.670 |
| | 80% AV | 94 | 1.09340 | 0.01167 | 1.915 |
| | 100% AV | 97 | 1.12040 | 0.01151 | 3.232 |

The summarized results for the 500 seconds simulation on Route 15 can be seen in Figure 9. The mobility increased with increasing AV penetration rates. However, the fuel economy benefit did not increase significantly between 35% and 65% AV penetration rates.

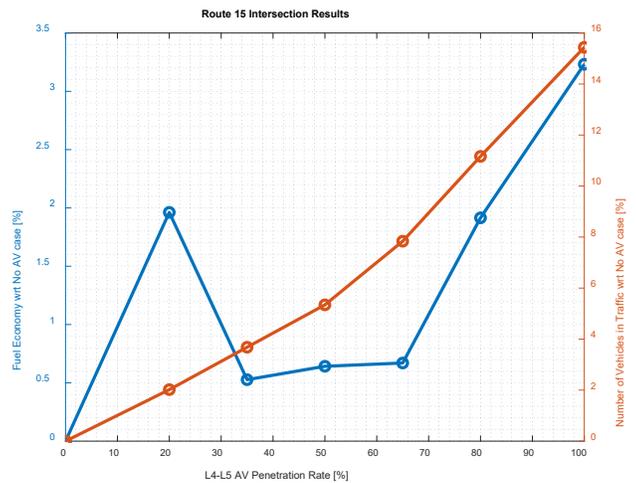

Figure 9: Fuel Economy and Number of Vehicles in Traffic Network vs L4-L5 AV Penetration Rate for an Intersection in Route 15

For the whole Route 15 and 3,600 sec simulation, the results are summarized and can be seen in Table 6. For Route 15, the overall trend was an increase in mobility and fuel economy benefit with increasing levels of L4-L5 AVs in the traffic network. The highest fuel economy benefit was reached by having 100% AVs in the network.

Table 6: Route 15, Full Route Results

| Route 15, Full Route Results – Urban Only Traffic | | | | | |
|---|---|---|---|---|---|
| Simulation Duration in seconds | AV Penetration Rate | Number of Vehicles | Total Fuel Consumption in US gallons | Fuel Consumption per Vehicle in US gallons | % Fuel Economy wrt no AV case |
| 3600 seconds | No AV | 1438 | 83.08310 | 0.05776 | 0.000 |
| | 20% AV | 1440 | 82.35840 | 0.05735 | 0.720 |
| | 35% AV | 1443 | 82.25670 | 0.05700 | 1.328 |
| | 50% AV | 1444 | 81.79750 | 0.05666 | 1.917 |
| | 65% AV | 1445 | 80.94580 | 0.05602 | 3.024 |
| | 80% AV | 1445 | 79.44190 | 0.05496 | 4.846 |
| | 100% AV | 1449 | 76.63170 | 0.05287 | 8.464 |

The summarized results for the 3600 seconds simulation on Route 15 can be seen in Figure 10. Both mobility and fuel economy improved with increasing AV penetration rate.

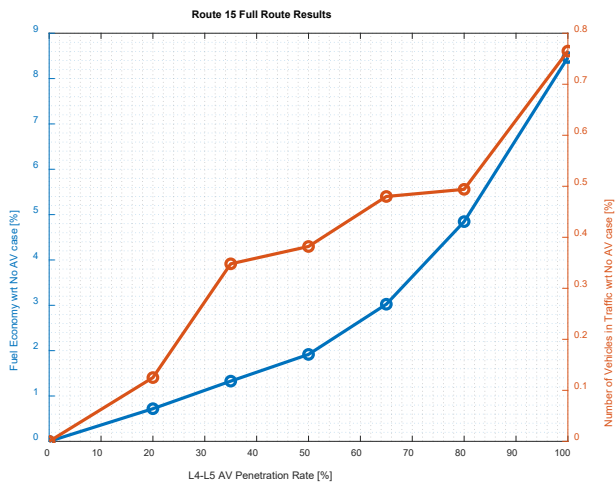

Figure 10: Fuel Economy and Number of Vehicles in Traffic Network vs L4-L5 AV Penetration Rate for Route 15

## *Traffic Network Route US 33*

For Route US 33, the merge in/out area studied can be seen in Figure 11. This merge in/out area is part of a whole traffic network that spans a freeway only route.

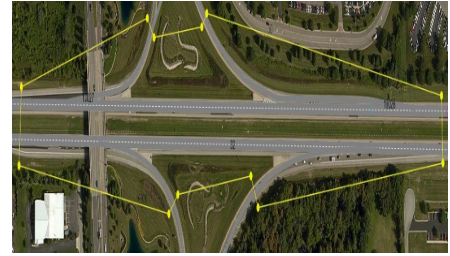

Figure 11: Route US 33 Merge In/Out Area in Vissim

Route US 33 can be seen in the map of Figure 12. It is the replica of a real route from Dublin to Marysville in Ohio. The total length of Route US 33 is 30.3 km.

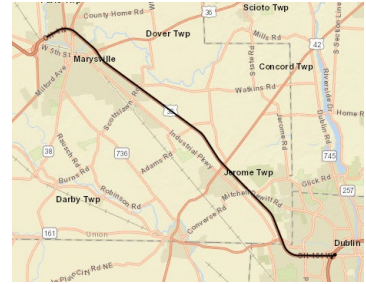

Figure 12: Route US 33

For the merge in/out area on Route US 33, the results can be seen in Table 7. The freeway portion of the route had 2 lanes, which enabled the AVs to change lanes frequently. The merge in/out lanes were single lanes. The mobility and fuel economy benefit increased by having higher levels of L4-L5 AVs in the network.

Table 7: Route US 33, Merge In/Out Results

| Route US 33, Merge In/Out Results – Freeway Only Traffic | | | | | |
|---|---|---|---|---|---|
| Simulation Duration in seconds | AV Penetration Rate | Number of Vehicles | Total Fuel Consumption in US gallons | Fuel Consumption per Vehicle in US gallons | % Fuel Economy wrt no AV case |
| 500 seconds | No AV | 177 | 3.54600 | 0.02007 | 0.000 |
| | 20% AV | 197 | 3.58380 | 0.01817 | 9.490 |
| | 35% AV | 198 | 3.61810 | 0.01824 | 9.123 |
| | 50% AV | 201 | 3.58130 | 0.01785 | 11.044 |
| | 65% AV | 204 | 3.57040 | 0.01754 | 12.591 |
| | 80% AV | 208 | 3.65840 | 0.01761 | 12.242 |
| | 100% AV | 215 | 3.71840 | 0.01731 | 13.769 |

For US 33, the merge in/out area results are summarized in Figure 13. It was observed that aggressive car following behavior with a small time gap is important for L4-L5 AVs, but their frequent and sudden lane changing behavior affects the mobility and fuel economy even more. Since the portion of the Route US 33 network studied here had 2 lanes, the AVs were free to change lanes frequently. This advantage of AVs resulted in an ever-increasing trend in both mobility and fuel economy benefit for the network.

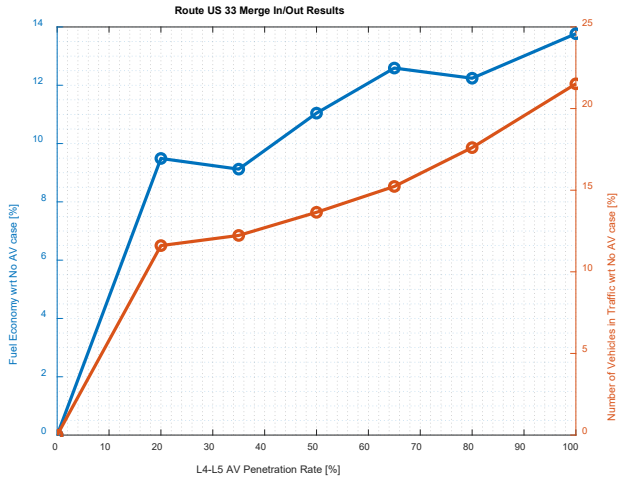

Figure 13: Fuel Economy and Number of Vehicles in Traffic Network vs L4-L5 AV Penetration Rate for a Merge In/Out Area in Route US 33

The results for the full Route 33 and 3,600 sec simulation can be seen in Table 8. The fuel economy benefit increased with increasing levels of L4-L5 AV penetrations.

Table 8: Route US 33, Full Route Results

| Route US 33, Full Route Results – Freeway Only Traffic | | | | | |
|---|---|---|---|---|---|
| Simulation Duration in seconds | AV Penetration Rate | Number of Vehicles | Total Fuel Consumption in US gallons | Fuel Consumption per Vehicle in US gallons | % Fuel Economy wrt no AV case |
| 3600 seconds | No AV | 2176 | 517.21540 | 0.23776 | 0.000 |
| | 20% AV | 2350 | 554.28280 | 0.23587 | 0.795 |
| | 35% AV | 2378 | 557.18610 | 0.23435 | 1.433 |
| | 50% AV | 2408 | 560.21790 | 0.23269 | 2.132 |
| | 65% AV | 2437 | 563.76760 | 0.23140 | 2.675 |
| | 80% AV | 2476 | 571.53420 | 0.23088 | 2.896 |
| | 100% AV | 2551 | 571.19560 | 0.22400 | 5.787 |

The summarized results for the 3,600 seconds simulation on Route US 33 can be seen in Figure 14. The mobility and fuel economy benefit increased with higher levels of L4-L5 AV penetration.

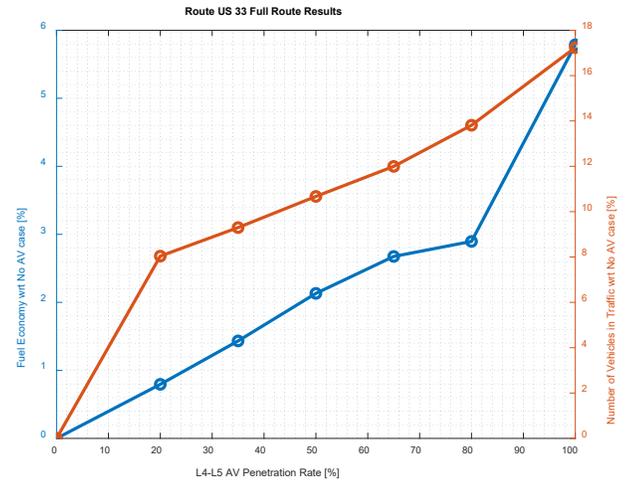

Figure 14: Fuel Economy and Number of Vehicles in Traffic Network vs L4-L5 AV Penetration Rate for Route US 33

### *Traffic Network Route COSI*

For Route COSI, the intersection studied can be seen in Figure 15. This intersection is part of a whole traffic network that spans an urban only route, the replica of a real route in Columbus, Ohio.

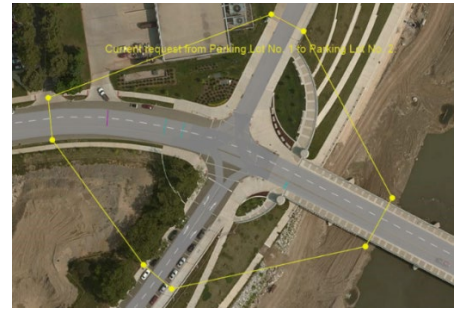

Figure 15: Route 15 Intersection in Vissim

Route COSI is also the route for a self-driving shuttle called Smart Circuit that is operating in downtown Columbus [14]. Route COSI can be seen in Figure 16 [15].

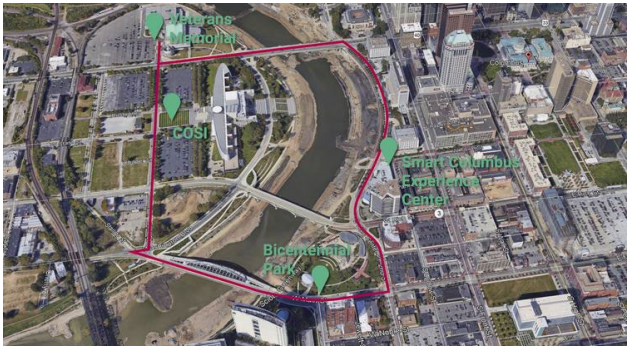

Figure 16: Route COSI [15]

For an intersection on Route COSI, the results can be seen in Table 9. From No AV to 20% AV case, the fuel economy benefit had a sudden jump. For the whole intersection, having AVs in the network benefited the mobility and fuel economy.

Table 9: Route COSI, Intersection Results

| Route COSI, Intersection Results – Urban Only Traffic | | | | | |
|---|---|---|---|---|---|
| Simulation Duration in seconds | AV Penetration Rate | Number of Vehicles | Total Fuel Consumption in US gallons | Fuel Consumption per Vehicle in US gallons | % Fuel Economy wrt no AV case |
| 500 seconds | No AV | 132 | 1.2904 | 0.009776 | 0.000 |
| | 20% AV | 135 | 1.2675 | 0.009389 | 3.957 |
| | 35% AV | 136 | 1.2714 | 0.009349 | 4.370 |
| | 50% AV | 137 | 1.25463 | 0.009158 | 6.320 |
| | 65% AV | 139 | 1.2792 | 0.009203 | 5.860 |
| | 80% AV | 138 | 1.24722 | 0.009038 | 7.549 |
| | 100% AV | 140 | 1.27023 | 0.009073 | 7.188 |

The summarized results for the 500 seconds simulation around an intersection in Route COSI can be seen in Figure 17. It is seen that having AVs improved the fuel economy around the intersection.

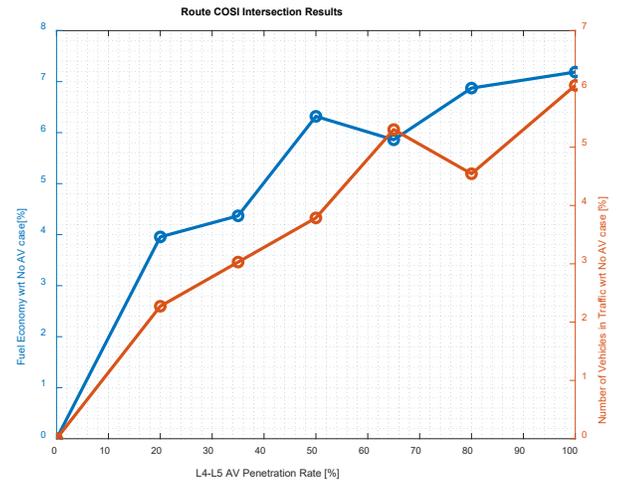

Figure 17: Fuel Economy and Number of Vehicles in Traffic Network vs L4-L5 AV Penetration Rate for an Intersection in Route COSI

The results for the 3,600 seconds full Route COSI simulation can be seen in Table 10. For the most part, increasing levels of AVs on Route COSI resulted in an increase in fuel economy benefit and mobility.

Table 10: Route COSI, Full Route Results

| Route COSI, Full Route Results – Urban Only Traffic | | | | | |
|---|---|---|---|---|---|
| Simulation Duration in seconds | AV Penetration Rate | Number of Vehicles | Total Fuel Consumption in US gallons | Fuel Consumption per Vehicle in US gallons | % Fuel Economy wrt no AV case |
| 3600 seconds | No AV | 2410 | 84.74760 | 0.03515 | 0.000 |
| | 20% AV | 2422 | 82.55580 | 0.03409 | 3.009 |
| | 35% AV | 2432 | 80.33030 | 0.03303 | 6.027 |
| | 50% AV | 2436 | 79.99210 | 0.03283 | 6.595 |
| | 65% AV | 2440 | 79.77000 | 0.03269 | 7.001 |
| | 80% AV | 2444 | 80.27640 | 0.03285 | 6.557 |
| | 100% AV | 2450 | 81.36700 | 0.03321 | 5.521 |

The summarized results for the 3,600 seconds simulation on Route COSI can be seen in Figure 18. The mobility of the whole network showed an increasing trend. The fuel economy benefit also showed an increasing trend, except between the 65% and 100% AV penetration rates. The reason for this trend could have been because of the capacity of the network. Some roads in Route COSI only had one lane. These single lane roads had queues forming during the simulations, resulting in longer queue lengths and lower fuel economy benefits.

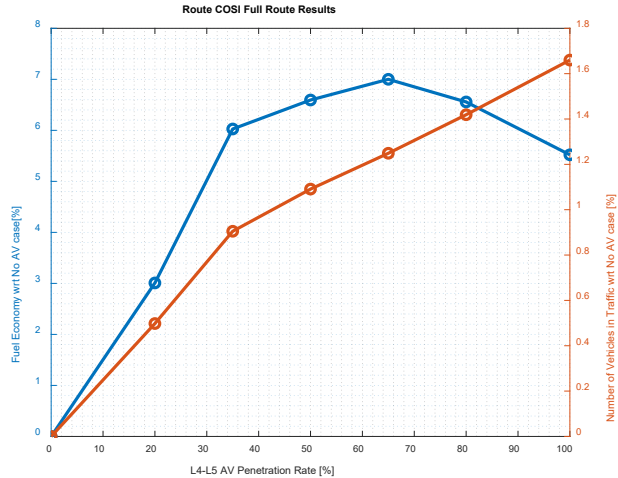

Figure 18: Fuel Economy and Number of Vehicles in Traffic Network vs L4-L5 AV Penetration Rate for Route COSI

## *Simulation Results for Various Metrics*

Simulation results regarding fuel economy benefit and mobility for each route have been presented individually in the previous sections. The following sections focus on different metrics to explain the effects of having varying levels of L4-L5 AVs in the traffic for all routes considered in this study.

The results of the 3,600 seconds full route simulations were used for each route. The metrics to be considered were fuel economy, mobility, average queue length, maximum queue length, average vehicle delay and average stopped delay.

### Fuel Economy

Using the node evaluation tool of Vissim, the total fuel consumed during the simulation was recorded. Then, the total amount of fuel consumed was divided by the total number of vehicles in the simulation to calculate the fuel consumption per vehicle in US gallons.

Fuel economy benefit with respect to the No AV case for each route were plotted together. The results are seen in Figure 19. Looking at the results, it can be seen that having L4-L5 AVs proved beneficial, i.e. positive, in terms of fuel economy at each penetration level for each route.

The existence of traffic lights and STOP signs on urban roads help the fuel economy with increasing number of L4-L5 AVs on roadways. Since AVs are smarter than conventional vehicles, they use information from infrastructure around them (V2I technology) to maintain their speed and eliminate unnecessary braking. This elimination of braking brings about fuel economy benefits with higher number of AVs on the roadways. For freeway driving, however, AVs cannot utilize their connectivity with the infrastructure. Conventional vehicles also show less braking on the highway, increasing their fuel economy, and diminishing the advantage AVs bring to the traffic network. That is why the lowest fuel economy benefit was from Route US 33 with freeway only driving in this study (Figure 19).

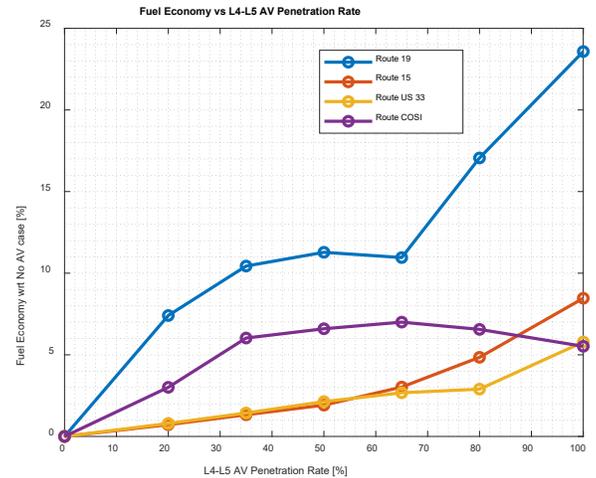

Figure 19: Fuel Economy vs L4-L5 AV Penetration Rate

### Mobility

For the mobility, the total number of vehicles in the simulation was considered. The increase in the number of vehicles in the whole traffic network was an increase in mobility.

Mobility benefit with respect to no AV case was plotted together for each route. The results are seen in Figure 20. Looking at the 4 routes considered in this study, for routes with only urban traffic such as Route 15 and Route COSI, a smaller increase in mobility was observed as the L4-L5 AV penetration rate increased.

Freeway driving requires a more deterministic control than the stochastic nature of urban driving. Since L4-L5 AVs behave deterministically, the mobility benefit of having a larger number of L4-L5 AVs in traffic increases more significantly on freeway only or mixed (urban + freeway) roads than urban only roads.

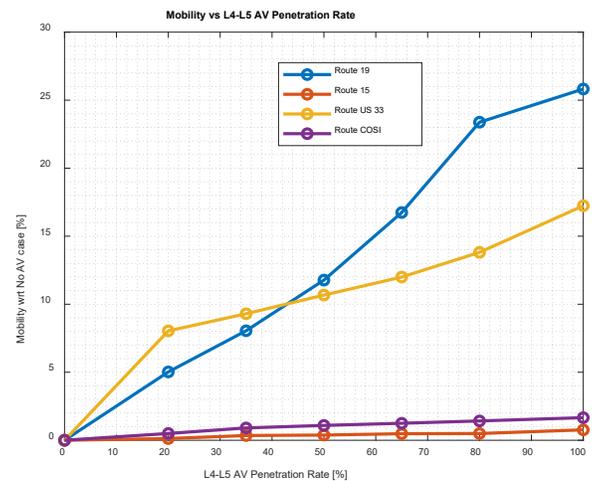

Figure 20: Mobility vs L4-L5 AV Penetration Rate

## Average Queue Length

The node evaluation tool in Vissim provides a performance metric called Average Queue Length. During the simulations, Vissim measures the current queue length at every time step and the arithmetic mean calculated is called the Average Queue Length. The changes in the Average Queue Length can also be changed by changing the waiting still time attributes of vehicles in Vissim. However, for this study, the waiting still time attribute of AV or non-AV vehicles were not modified.

The results are shown in Figure 21. For each route, having an increased number of L4-L5 AVs resulted in smaller average queue length as compared to the non-AV case (negative values).

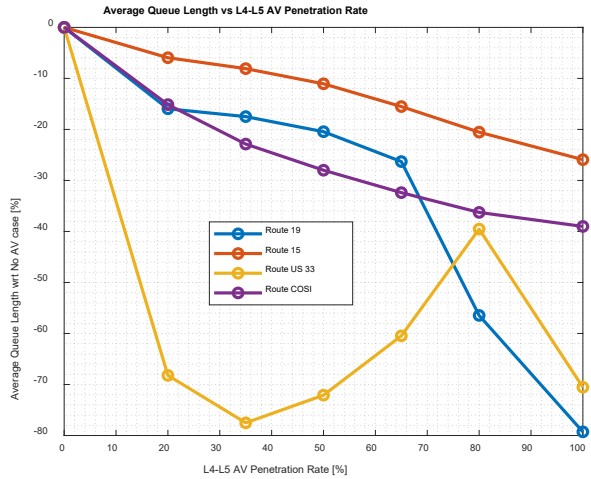

Figure 21: Average Queue Length vs L4-L5 AV Penetration Rate

## Maximum Queue Length

The node evaluation tool in Vissim provides a performance metric called Maximum Queue Length. During the simulation, Vissim measures the current queue length at every time step and maximum queue length is recorded.

The results are shown in Figure 22. Looking at the results, it is seen that the maximum queue length generally decreased as the L4-L5 AV penetration rate increased.

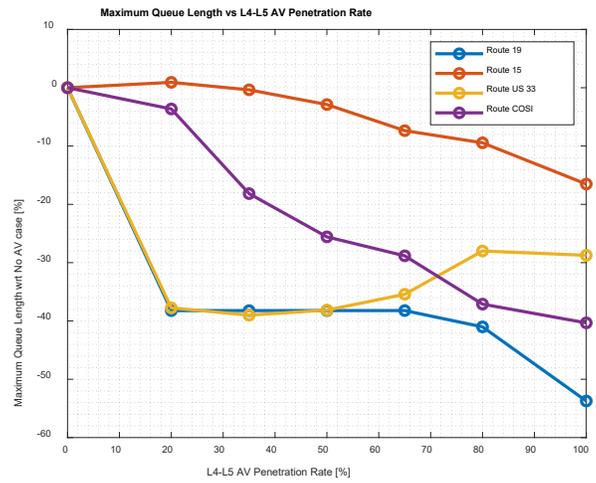

Figure 22: Maximum Queue Length vs L4-L5 AV Penetration Rate

## Average Vehicle Delay

The node evaluation tool in Vissim provides a performance metric called Average Vehicle Delay. During the Vissim simulation, the theoretical ideal travel time for each vehicle was calculated. Theoretical travel time is the travel time that could be achieved by the vehicle if there were no other vehicles around it (no traffic, single vehicle only) and no signal controls on the traffic lights (assumes each traffic lights was green). Then, the theoretical travel time was subtracted from the actual travel time for each vehicle, and the average for all vehicles in the simulation was calculated in seconds.

The results are shown in Figure 23. The general trend shows that having an increased number of L4-L5 AVs in the traffic was beneficial in reducing average vehicle delay due to the communication capabilities of the AVs with each other and being aware of other vehicles around them except route US 33 up to 70% penetration rate.

Route US 33 only had freeway traffic, meaning that there were not STOP signs or traffic lights on the road. The non-AVs on Route US

33 also travelled at high speed without much travel delay, resulting in a delayed benefit with increasing AV penetration rates.

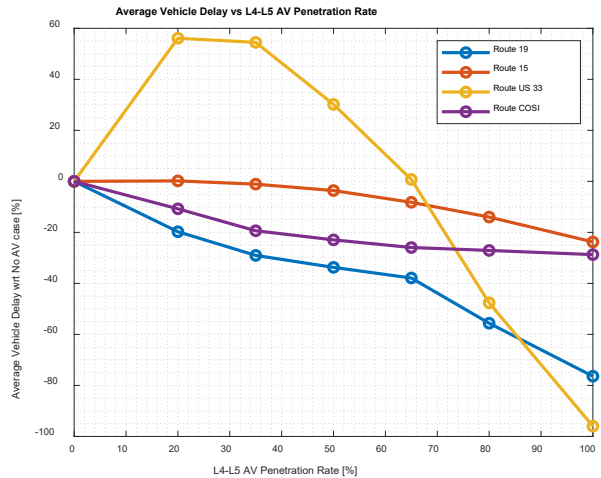

Figure 23: Average Vehicle Delay vs L4-L5 AV Penetration Rate

**Average Stopped Delay**

The node evaluation tool in Vissim provides a performance metric called Average Stopped Delay. During the simulation, how long each vehicle stopped was recorded. Then, the total stop duration was divided by the total number of vehicles in the simulation and Average Stopped Delay in seconds was found.

The results shown in Figure 24 demonstrate that average stopped delay was less for all AV penetration rates for all four routes as compared to the non-AV case.

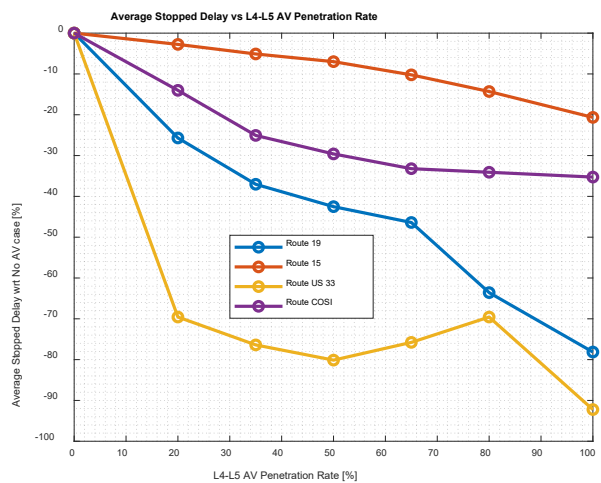

Figure 24: Average Stopped Delay vs L4-L5 AV Penetration Rate

**Emissions CO, NOx and VOC**

The node evaluation tool in Vissim provides a performance metric in terms of vehicle Emissions. the abbreviations in the title are as follows: CO stands for Carbon Monoxide, NOx stands for different Nitrogen Oxide compounds and VOC stands for Volatile Organic Compounds.

For CO, NOx and VOC, total quantity of each emission type was recorded during the simulation in grams. Then, the total emission amount was divided by the total number of vehicles in the simulation to get the average emissions.

The results are shown in Figure 25. The general trend shows that having AVs in the traffic reduced the emissions for each route.

The most drastic change was seen in the Route 19 mixed route. Between AV penetration rates of 65% and 80%, the emissions reduced drastically.

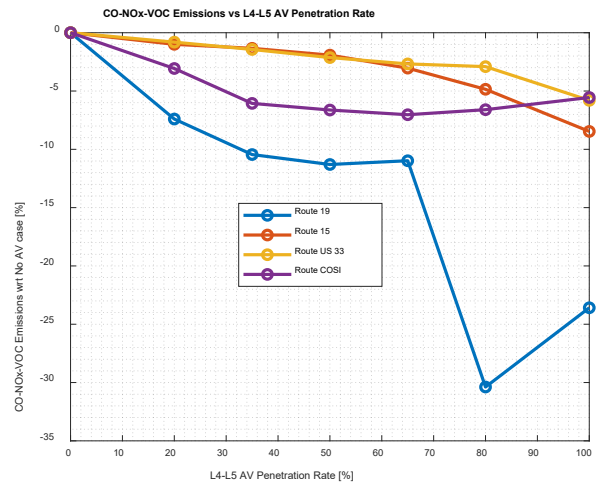

Figure 25: CO-NOx-VOC Emissions vs L4-L5 AV Penetration Rate

## Conclusions and Future Work

In this paper, varying levels of L4-L5 AVs were simulated in four different traffic networks, which had characteristics of urban only traffic, freeway only traffic and mixed traffic conditions. The effects of having varying numbers of L4-L5 AVs in the network were studied in terms of fuel economy, mobility, emissions and queue length.

The results show that while increasing penetration rates of L4-L5 AVs generally improve overall fuel efficiency and mobility of the traffic network, there were also cases when the opposite trend was observed. Having varying levels of L4-L5 AVs in traffic also generally helps in decreasing the queue lengths, vehicle emissions and vehicle delays.

The way that the AVs increase the mobility and fuel economy are by sudden lane change maneuvers they perform when the preceding vehicle they encounter is traveling at a slower speed than their own speed. Another observation drawn from the AVs during simulations is that they do cooperative lane changing. These observations are significant as they also provide a basis for how AV controls should be designed for overall fuel efficiency and mobility improvements.

For future work, fuel optimal models can be incorporated into Vissim. By doing so, the effects of having L4-L5 AVs with fuel optimal velocity profiles can be examined. One such fuel optimal model which is a connected vehicle application is Pass-at-Green [16] that adjusts AV speed to pass at the green light and can be integrated into Vissim [17].

Another direction that could be taken to study AVs in traffic could be done by modeling larger networks with more intersections. Pedestrians can also be added to the simulations to see how the AV performance changes with pedestrian to vehicle (P2V) interaction.

For the real-world application of this work, if real world data could be collected from experimental vehicles in controlled testing grounds or from public road deployments, the AV behavior could be modeled more accurately. Then, those models can be tested in simulation environments for various scenarios and the results can be used to improve the capability of AVs.

The traffic and AV traffic simulation methods introduced in this paper can be incorporated and used in developing, simulating and evaluating methods as diverse as controls, automotive controls including powertrain control, advanced driver assistance systems, connected vehicles and autonomous driving systems [19-73].